# Charge Transfer in Classical Molecular Dynamics Simulations of Met-enkephalin: Improving Traditional Force Field with Data Driven Models

Tiange Dong[1], Fang Liu[1], Likai Du[1]*, Dongju Zhang[2], Jun Gao[1]*

[1]Hubei Key Laboratory of Agricultural Bioinformatics, College of Informatics, Huazhong Agricultural University, Wuhan, 430070, P. R. China

[2]Institute of Theoretical Chemistry, Shandong University, Jinan, 250100, P. R. China

*To whom correspondence should be addressed.

Likai Du: dulikai@mail.hzau.edu.cn; Jun Gao: gaojun@ mail.hzau.edu.cn

**Abstract:**

The charge transfer and polarization effects are important components in the molecular mechanism description of bio-molecules. Classical force field with fixed point charge cannot take into the account of the non-negligible correlation between atomic charge and structure changes. In this work, high throughput *ab initio* calculations for the pentapeptide Met-enkephalin (MetEnk) reveal that geometric dependent charge transfer among residues is significant among tens of thousands of conformations. And we suggest a data driven model with machine learning algorithms to solve the geometric dependent charge fluctuations problem. This data driven model can directly provide *ab initio* level atomic charges of any structure for MetEnk, and avoids self-consistent iteration in polarizable force field. Molecular dynamics simulations demonstrated that the data driven model provides a possible choice to describe the explicit charge flux with minor modification of available classical force fields. This work provides us an alternative molecular mechanism model for future dynamics simulation of oligopeptides.

**Introduction**

Molecular dynamics (MD) methods are widely used in the studies of bio-molecular systems. The predictive quality of MD simulations depends on the accuracy of the potential energy surface (PES) and its ability to sample the effective configurations or phase space. As a long-range inter-molecular interaction, electrostatic term plays an important role to rationalize the problems of the bio-molecular folding and functions. At present, the majority of MD simulations are performed with non-polarizable (fixed charge) force fields, such as AMBER[1], CHARMM[2], GROMOS[3] and OPLS-AA[4]. Their empirical function formulas and parameter sets only implicitly incorporate charge fluctuation effects (i.e. charge transfer and polarization) in an averaged way.

It is well-known that the atomic charges depend on molecular conformations, and the charge fluctuations are usually believed to be non-negligible in metal ion or charged molecular systems.[5-6] Jensen and co-workers have revealed the importance of the charge flux in force field modeling for peptide models of the 20 neutral amino acids.[7] Thus, it is essential to explicitly take into account of the polarization and charge transfer contributions in potential energy functions.[8-12]

Over the recent years, the development of advanced force fields incorporating charge transfer and polarization components has been reported for many systems ranging from liquid water to metalloenzymes.[8-11, 13-15] For example, the AMBER and CHARMM force fields have been extended to include polarization, with atomic dipole polarizability[16-18] and the Drude particle model[19]. The AMOEBA force field[20] is one example of molecular mechanism model with advanced function formulas, which explicitly treats polarization effects in various chemical and physical environments. The explicit polarization (X-Pol) method represents recent efforts to construct the inter-atomic potential explicitly from quantum mechanics[21-22]. The partial polarizable force field, such as the effective polarizable bond (EPB) model[23] is also developed to take into account the effective atomic charges fluctuation according to the local environment. These featured force fields offer a clear and systematic improvement in functional forms involving various kinds of coupled polarizable sites.[24-26]

It appears a good choice to allow the atomic charges to depend on the atomic positions, but an efficient parametrization of this geometry dependence is an open question. Recently, the data intensive or data driven paradigm, popularly referred to as machine learning (ML), began to emerge as viable pathways for the creation of inter-atomic potentials by training against huge *ab*

*initio* level data sets.[27-32]

As a practical issue, the prospect of using ML algorithms to tackle the flood of dynamics data to yield statistical significance is indeed very promising. Using ML algorithms to reproduce *ab initio* calculation results would greatly reduce the computational cost without loss of the accuracy.[33] The ML algorithms have been applied to predict PES at QM and QM/MM level successfully using neural networks (NN) model,[34-35] or accelerate the ab initio MD simulation of material systems.[36-37] The data driven model with ML algorithms have also shown its abilities to automatically probe the properties of dynamics data, parameter fitting and improve the quality of electronic structure methods.[38-42] Therefore, it is very necessary to devote our efforts to the data driven force field ($D^2FF$) scheme and improve the polarization and charge transfer descriptions in the traditional force fields.

In this work, we explore the conformation dependent atomic charges with the electronic structure methods for the pentapeptide Met-enkephalin (MetEnk, YGGFM), which is one of the smallest neurotransmitter peptides.[43] By using this data set as the starting point, the nonlinear relationship between atomic configurations and atomic charge could be derived. Then, the idea of $D^2FF$ scheme is proposed with ML techniques. And the atomic charges of the MetEnk are directly estimated with our approach of prediction with ensemble models (PEM) on the finite geometric patterns. As pointed out in our previous work[31, 44], the PEM algorithm could correctly predict the point charge contribution with no need of any training data except the clustering analysis. Beyond developing more dedicated potential function forms, we adopt the available function forms of traditional force fields, such as AMBER and CHARMM in our studies. The classical MD simulations were performed for the neutral and charged MetEnk with the $D^2FF$/PEM scheme on the top of AMBER force field (namely, PEM@AMBER), and compared with AMBER *ff14SB* force field. In this fashion, the charge flux among residues along the MD trajectories can be explicitly observed in the PEM@AMBER scheme.

## 2. Theory and Computational Methods

### 2.1 Big Data Sets from MD Simulations

The first step to develop any data driven model is the data collection procedure. Then, available ML algorithms can be readily applied for desired purposes. Here, the initial data sets were obtained from MD simulation itself with the popular AMBER force field. The Amber *ff*14SB force field parameters[45] were adopted in the MD simulations. The geometries of the neutral and charged MetEnk have been prepared by the *tleap* module of AmberTools package[46]. The neutral MetEnk system was terminated by ACE and NME groups.

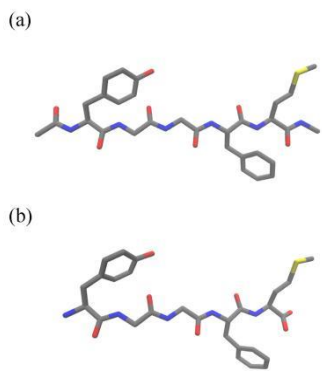

**Figure 1.** The initial structures of neutral (a) and charged (b) MetEnk molecules are shown. The hydrogen atom is not shown for clarify. The torsional angles involving backbone atoms are used as molecular descriptors for clustering analysis.

The initial structure of MetEnk was set to be in a linear conformation (Figure 1). Both oligopeptide models were subjected to identical energy minimization procedures to remove the possible overlapping. The minimization included 100 steps of steepest descent, followed by 500 steps of the conjugate gradient method.

Following the minimization, subsequent MD simulations were performed both in *vacuo* and the Generalized Born implicit solvent (GBIS) model. The NVT ensemble was used in the MD simulation with a constant temperature of 300 K. The integration time-step was set to be 1.0 fs and Langevin thermostat was used for temperature control. For each system, we performed 50 dynamics trajectories (~1.0 ns) with random initial velocities, which seemed to be sufficient to observe the most signficiant folding pathway of the pentapeptide MetEnk[47]. The structures of local energy minima along with the MD trajectories were taken as our data sets, namely 15778 and 18556 snapshots for neutral and charege MetEnk systems. All MD simulations were performed using NAMD package[48].

### 2.2 Clustering Analysis

To characterize the structural diversity of the geometric data sets, we used K-means clustering to classify the conformation space into a set of discrete patterns or clusters. Clustering the data into different groups makes the analysis of ensembles of untractable sizes more amenable. As a simple and robust algorithm, the K-means is very suitable to identify the most significant structural features. The K-means clustering algorithm aims to partition *n* observations into *k* clusters, in which each observation belongs to the cluster with the nearest mean, serving as a prototype of the cluster.

An arbitrarily low number of clusters may lead to important information being missed, while too many clusters would be unnecessarily redundant. Thus, it is necessary to set an appropriate number of clusters *a priori*. We have tested various cluster numbers (i.e. $k = 2\sim100$), and 8 and 16 clusters was sufficient and efficient for our subsequent analysis of neutral and charged systems. The molecular internal coordinates, namely torsional angles involving backbone atoms, were used as molecular descriptors for clustering (Figure 1). This protocol has been implemented with scikit-learn module[49] in Python.

**2.3 The Data Driven Charge Fluctuation Model**

The general function form of the traditional force field[1] was retained in our data driven model. The electrostatic energy was still described by Coulomb interactions of pairwise point charges. And we attempted to directly update the conformation dependent atomic charges in this fixed point charge force field. Such treatment can take into account of charge flux effects, and the modification of the available force field was minor. The extensive QM calculations of such huge data sets were automatically done on a distributed computer cluster with a series of Python scripts interface with quantum chemistry package. The function form is given as follows,

$$U(\mathbf{R}) = \sum_{bonds} k_b(b-b_0)^2 + \sum_{angles} k_\theta(b-b_\theta)^2 + \sum_{torsions} \frac{V_n}{2}[1+\cos(n\phi-\delta)] \\ + \sum_{improper} V_{imp} + \sum_{vdw} 4\varepsilon ij\left(\frac{\sigma_{ij}^{12}}{r_{ij}^{12}} - \frac{\sigma_{ij}^{6}}{r_{ij}^{6}}\right) + \sum_{elec} \frac{Q_i(\mathbf{X})Q_j(\mathbf{X})}{r_{ij}} \quad (1)$$

As shown in Eq. (1), the electrostatic interaction was improved to be geometric dependent, for which the point charge was a function of the collective torsional angles (**X**) involving backbone atoms.[7] This was achieved by training the *ab initio* level atomic charges for various conformations from clustering analysis, and then the conformation similarity was explored to estimate the atomic charges through the evolution of MD simulations.

In previous work, we have proved that the conformations in the same cluster should have much similar properties (low bias) than among different clusters (high variance).[31] And the molecular properties, i.e. atomic charges, could be represented by the weighted combination of only finite geometric patterns or clusters.

An explicit and efficient description of the geometry dependence of atomic charges at force field level is still an open question. Here, we only provided a general version of ensemble averaging algorithm, namely Prediction with Ensemble Models (PEM) algorithm, to build a model and predict reliable molecular properties. Note that, the multiple conformations are nowadays commonly used in the partial charge fitting scheme.[50-51] If the molecule descriptors ($\mathbf{X}$), (namely torsional angles involving backbone atoms) were specified, the estimated molecular property $Q$ (namely atom charges) can be defined as

$$Q(\mathbf{X}) = \sum_{i=1}^{M} \sum_{j=1}^{n} \omega_{ij}(\mathbf{X}) \mathbf{T}_{ij}(\mathbf{X}) \tag{2}$$

In Eq. (2), $\mathbf{T}_{ij}$ is the atomic charge vector of a specific structure ($j$) in one cluster ($i$), and the distinct clusters are generated by the K-means algorithm. And $M$ is the number of clusters, $n$ is the number of possible structures in each cluster, and ω is a set of weights. The kernel-based ML methods[31, 52] is adopted to directly obtain the weights (ω), which tends to be somewhat easier to set up in practice than the artificial neural networks.

The kernel function should be continuous in the input space of molecular descriptors, so any small perturbation of the system does not change the results too much, and the $\mathbf{X}$ far from a specific structure should have less weight, because such structures have little similarity on the conformation space. Here, the weight/kernel function is defined as a function of general distances between the molecular descriptors belong to an arbitrary geometry and a set of known geometries.

$$\omega_{ij} = \frac{1/u_{ij}}{\sum_{i=1}^{N} (1/u_{ij})^4} \tag{3}$$

In Eq. (3), N is the number of known molecular geometries, and $u_{ij}$ is the general distance, which can be defined as

$$u_{i\,j}(\mathbf{X}) = \sqrt{\sum_{k=1}^{p}(\mathbf{X}-\mathbf{X}_k')^2} \qquad (4)$$

In Eq. (4), $p$ is the number of molecular descriptors (internal coordinates), and $\mathbf{X}_k'$ is known values of molecular descriptors, namely backbone torsional angles in this work.

The PEM algorithm is used to replace the fixed point charge model in AMBER force field, in order to predict the geometric dependent charge fluctuations. One possible problem with our current approach is the determination of how far we can move away from the data set in geometric space before the estimation approach fails. The general distance can be used as a criterion to determine the quality of the prediction. If the weight for any geometric cluster is smaller than 10%, the averaged atomic charge from the entire geometric clusters will be used. The PEM algorithm may avoid the problem of polarization catastrophe at short distances[53-54] and also self-consistent calculating the atomic charge distribution at each MD step in most available polarizable force field.

**2.4 Implementation Details**

The effective atomic charges population for each conformation in our data sets were obtained by fitting the gas-phase electrostatic potential calculated by HF/6-31G* using RESP, as the protocol in Amber GAFF model.[55] The geometric dependence of atomic charges was analyzed statistically. The extensive QM calculations of such huge data sets were automatically done on a distributed computer cluster with a series of Python scripts interfaced with Gaussian 09 package[56].

The PEM algorithm requires the number (M) of geometric clusters for prediction (see Eq. 1), which is set to be the total number of geometric clusters. For the MetEnk model system, there is only a very finite set of distinct patterns (i.e. M=8 or 16). Since the general distance is inversely proportional to the contribution of the molecular properties, one can optionally reduce this number, if the number of distinct patterns is very large (i.e. a few thousands).

The number ($n$) of possible structures in each pattern is another critical parameter to be determined. We use the "mini-batch" option to reduce the total number of numerical calculations, and also overcome the defects of stochastic behavior.[31]

The PEM algorithm typically scales as O (M·$n$), whereas M and $n$ are very small constant value in the "mini-batch" option. In summary, the "mini-batch" option is used with M=8/16 and $n$=10 for the neutral and charged systems. This means that all of meta-stable patterns (8/16clusters) were used to estimate the molecular properties of any unknown geometry, meanwhile, we randomly select ten geometries (*n*=10) from each pattern to build parameter sets. Once constructed, these data sets do not need to contain millions of structures (or more) for prediction. In short, we can use PEM algorithm to convert big data into smart data.

The $D^2$FF/PEM scheme has been integrated as a plugin into NAMD package. For the modified MD package, the geometric dependent atomic charges would be updated by $D^2$FF/PEM plugin at each time step, through the evolution of MD trajectories.

And the electrostatic interactions from fixed point charge model were replaced with the new atomic charges. Undoubtedly, such minor modification to the AMBER force field can be easily implemented in any available MD package. This is quite similar as the space domain electrostatic embedding scheme between the QM and MM regions in QM/MM method, and the $D^2$FF/PEM improve the electrostatic interaction by incorporating the available trajectory information in the time domain.

In $D^2$FF/PEM scheme, the reliable electrostatic potential energy and its gradients can be obtained with minor extra calculations. We have extended the PEM algorithm on the basis of AMBER *ff14SB* force field, namely, PEM@AMBER model. The same MD simulation protocol is adopted in this work. The workflow of $D^2$FF/PEM scheme is given in Figure 2.

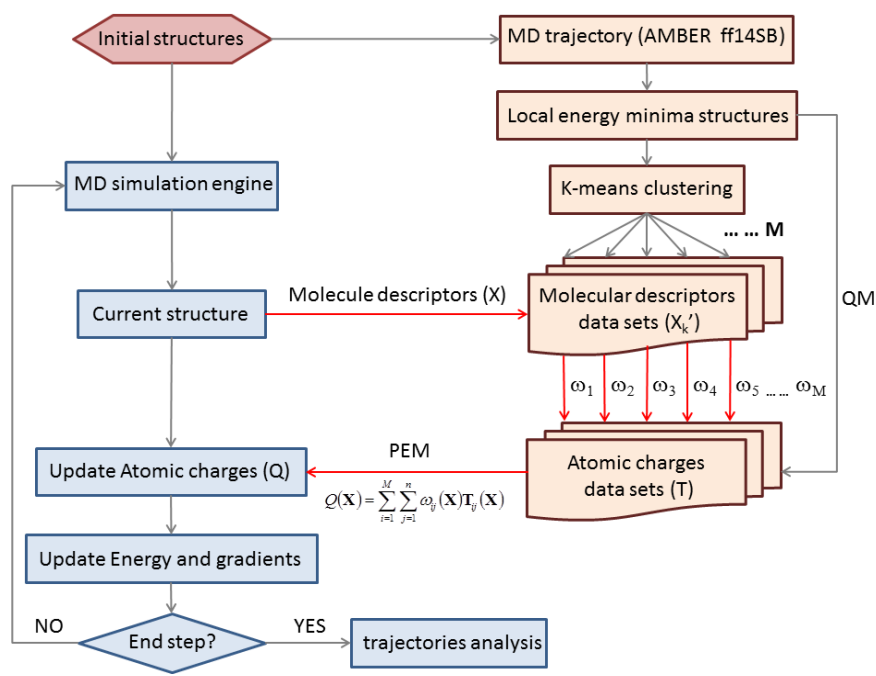

**Figure 2.** A schematic illustration of the $D^2$FF/PEM workflow.

**Results**

First, the clustering algorithm was applied to automatically partition the data sets into geometrically distinct clusters, which provided a clear-cut representation of the possible dynamics state of MetEnk. Figure 3 shows the overlapped molecular geometries for each cluster. Some attempts have been performed to vary the number ($k$) of distinct clusters to reach better structural similarity within each cluster. And the number of $k = 8$ and 16 is used for neutral and charged systems. The torsional degrees of freedom involving backbone atoms were considered in the clustering analysis, which has been suggested to be the main contribution to conformational changes of atomic charges and molecular dipole moments.[7] This also show good agreements with dihedral clustering base on Markov state model (MSM)[47]. The clustering results clearly reveal the possible folding pathways of MetEnk from meta-stable linear conformations to the stable folded conformations.

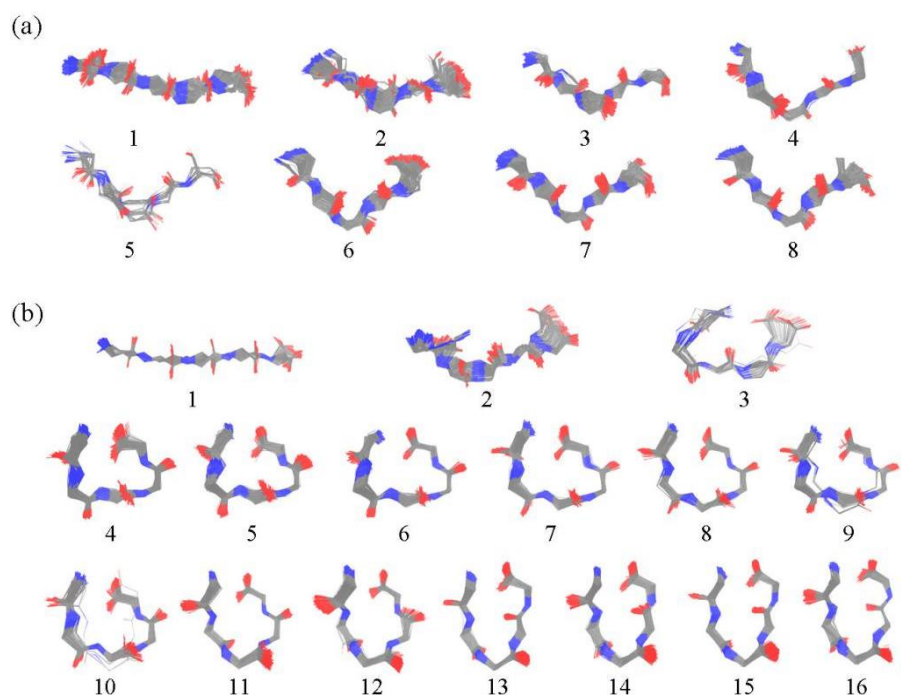

**Figure 3.** The intrinsic patterns of the neutral (a) and charged (b) MetEnk system from clustering analysis are given. The hydrogen atoms are not displayed for clarify.

To emphasize the necessary of atomic charges variation depending on the geometric changes, we try to understand the dynamic dependence of the atomic charge fluctuations in the existing non-polarizable models.[57-59] Here, the atomic charge population was analyzed for each structure in our data sets. Figure 4 shows the charge distribution at atomic level and residue level, for both neutral and charged systems. Significant variation of the charge population is observed for each residue. This is especially true for the charged MetEnk, with largest standard deviation (σ) of nearly ~0.5 e. Moreover, the charge population at atom level varies even larger. For a few atoms, the charge fluctuation is approximately ~0.9 e. The significant charge fluctuation is especially true for the charged MetEnk system. The point charge variation is suggested to be very important in molecular dynamics of bio-molecules.[7, 60] These statistical results suggest that it is essential to include polarization and the charge transfer effects in MD simulations.

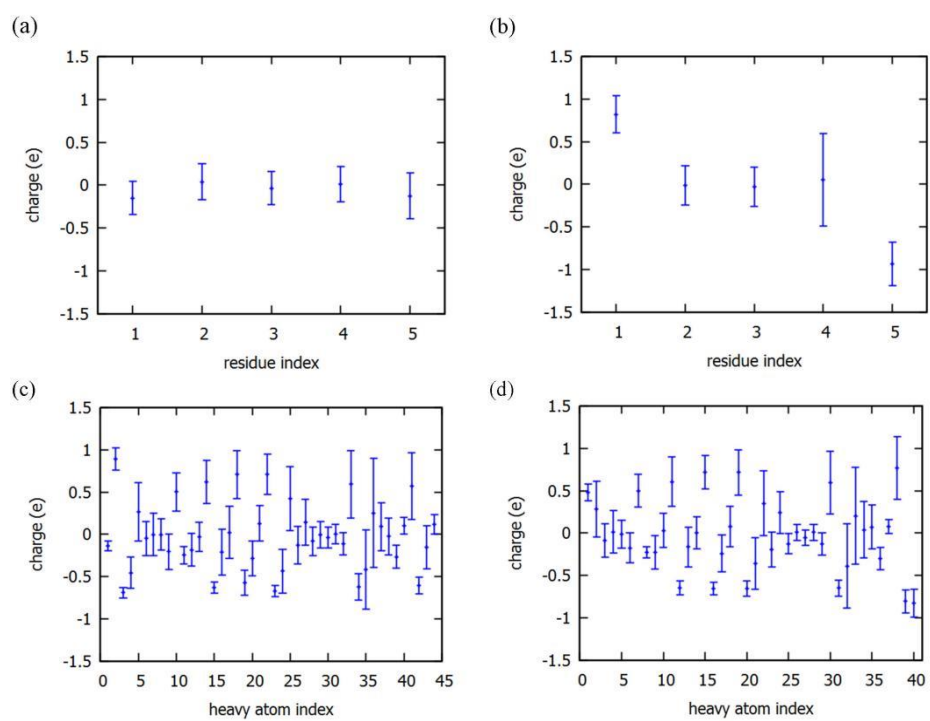

**Figure 4.** The residue level charge distribution of the neutral (a) and charged (b) MetEnk systems, and the atomic level charge distribution of the neutral (c) and charged (d) MetEnk systems. Only the atomic charges of heavy atoms are shown.

Next, the MD simulations were performed with the $D^2FF/PEM$ scheme, which provided a powerful way to update atomic charges along MD trajectories. Figure 5 shows the residue level charge variations along the time series of MD trajectories, which were performed with PEM@AMBER force field. Interestingly, the charge fluctuation at the residue level can be directly observed during the MD trajectories. The averaged residue charges population (Figure 5a and 5b) tends to be converged to be constants with minor fluctuations after the folding process finishes. This indicates the current fixed point charge model is quite reasonable for the MD simulation of equilibrium folded states, which ensures the success of traditional force fields for ensemble averaging properties. The residue charge is not purely an integer as in most fixed point charge model ($\delta=\pm0.1e$). This is especially true for any typical trajectory (Figure 5c and 5d). There were several time points with significant charge fluctuation for a typical trajectory, which can be corresponded to larger structural fluctuations as discussed below.

Note that, it is also very important to determine the uncertainty in the prediction, and we have evaluated the confidence of the results by comparing with ab initio calculations of the conformations randomly sampled from MD trajectories. Generally, the available representative sets of molecular conformations can be used in a nonlinear combination to describe any other structure required in poly-dispersed ensembles. Therefore, the PEM@AMBER force field provides us a possible choice to understand the charge flux among residues for MetEnk.

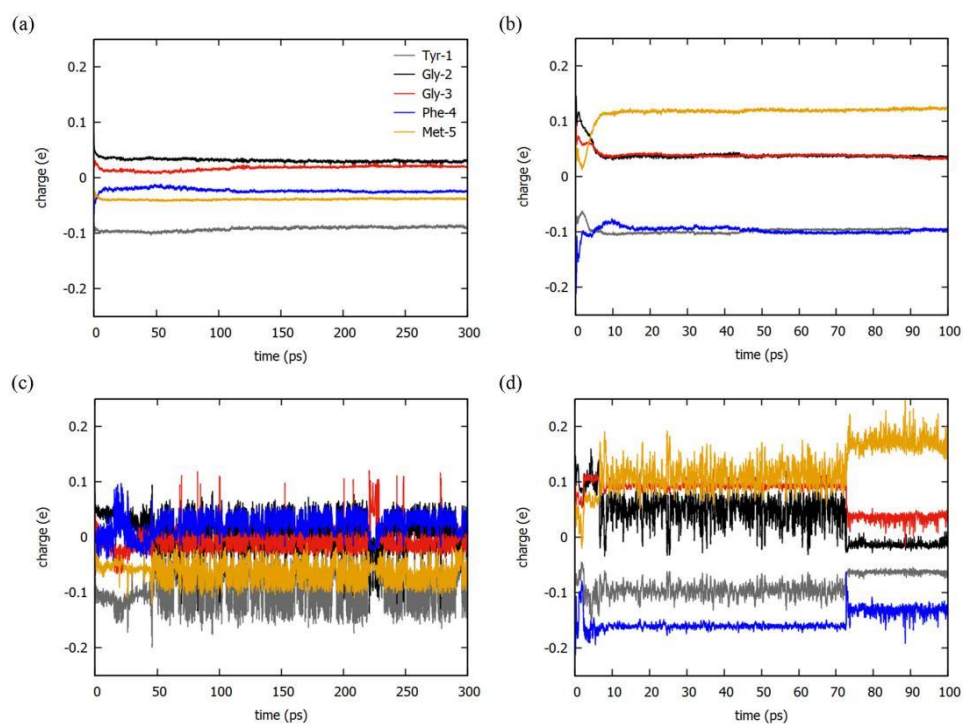

**Figure 5**. The time evolution of residue charges population averaged over MD trajectories, for the neutral (a) and charged (b) systems. The residue charges population for a specific trajectory is also given, for the neutral (c) and charged (d) systems. Note, the residue charge of Tyr-1 and Met-5 for the charged system is shifted to near zero by plus ±1.0 e.

In Figure 6, the MetEnk molecule is colored to represent the atomic charges fluctuation in contrast to the atomic charges of initial structures. The blue color represents negative fluctuation, while the red color refers to the positive fluctuation from the initial structure. The color depth represents the numerical value of atomic charge fluctuations. The time evolution of atomic charges significantly varied during the peptide folding process. Obviously, the atomic charge flux is possible through the entire MD trajectories. In contrast to the neutral system, the geometric dependent atomic charge flux is more obvious for the charged MetEnk system. Therefore, the charge flux becomes important during peptide folding and may contribute to the details of inter-atomic interactions.

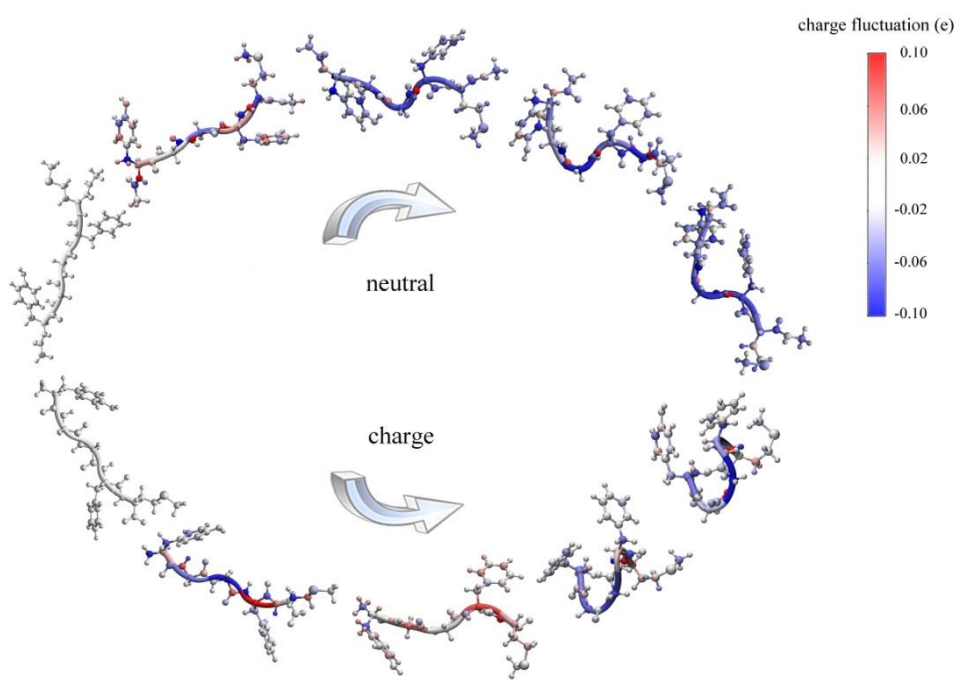

**Figure 6.** The time evolution of atomic charges fluctuation is shown for specific trajectories of the neutral and charged MetEnk systems. The color depth represents the numerical value of atomic charges fluctuation with respect to its initial structure.

Then, we pay our attention to the structural changes along the folding processes. The MD simulations were performed with 50 trajectories for both netural and charged MetEnk systems.

Figure 7 shows the time evolution of root mean square deviation (RMSD) averaged over 50 trajectories, performed with AMBER and PEM@AMBER force fields. The charge fluctuation PEM@AMBER model could reach equilibrium faster. This is consistent with several works with polarization force fields, which accelerate the protein folding processes.[61-64] Generally, the MetEnk could reach its folding state within hundred of picoseconds, starting from a linear conformation. The folding rate of the pentapeptide MetEnk obviously speeds up, using the PEM@AMBER model. For the neutral system, the following oscillation time scale to reach equilibrium is similar for both force fields, i.e. 150~200 ps. Meanwhile, the folding time scale of charged MetEnk is relatively modified by the PEM@AMBER model (20 ps *v.s.* 50 ps).

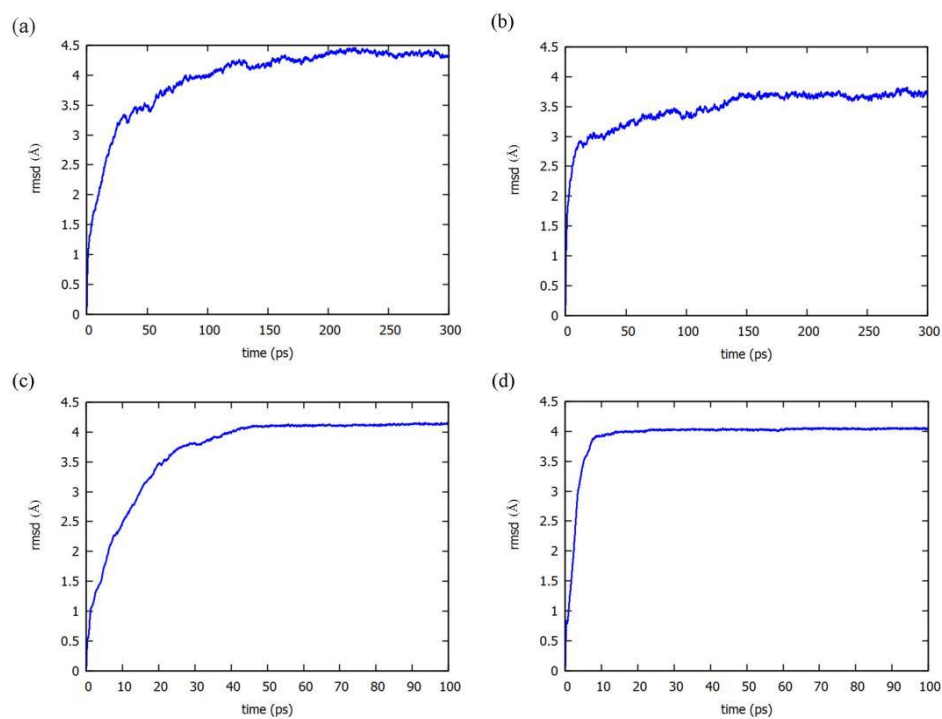

**Figure 7.** The RMSD for the neutral and charge MetEnk systems, averged over 50 trajectories. (a) the neutral system with AMBER; (b) the neutral system with PEM@AMBER; (c) the charged system with AMBER; (d) the charged system with PEM@AMBER.

The structural features of the charged MetEnk system were also studied for a typical MD trajectory (Figure 8). Two regions with large RMSD fluctuation are identified (Region 1 and 2) in Figure 8a, and the overlapped peptide backbone is also given to illustrate their global structural similarity. Note that, the backbone color refers to the charge fluctuations with respect to its initial value. Indeed, the significant conformation motions are coupled with obvious charge flux effects. As shown in Figure 8b, the residue charge population shows sharp fluctuation at the same time of structural changes. This also reflects the possible charge transfer event among residues. For example, the residue charges of Gly-2, Gly-3 and Phe-4 drop to near zero, while the dominated electrostatic interactions from the residue Tyr-1 and Met-5 are enhanced, at the timescale of 70~80 ps. Thus, the charge flux effect should modify the peptide folding process, especially for the charged MetEnk system.

Figure 8c shows further structure details about hydrogen bonds, which reveal remarkably structural change along with the charge flux events. At the Region **1**, the hydrogen bond of carboxyl oxygen of Met-5 with Tyr-1 changes to the backbone amide hydrogen of Gly-3. At the Region 2, the number of hydrogen bonds increases to three, for which the carboxyl oxygen atom of Met-5 forms extra hydrogen bonds with of Tyr-1 and Gly-3. Such variation of hydrogen bonds is not commonly observed in the dynamic trajectories with AMBER force field, starting from the same initial conditions. Therefore, the charge fluctuation in PEM@AMBER force field introduces a few flexible hydrogen bond variations at atomic resolution.

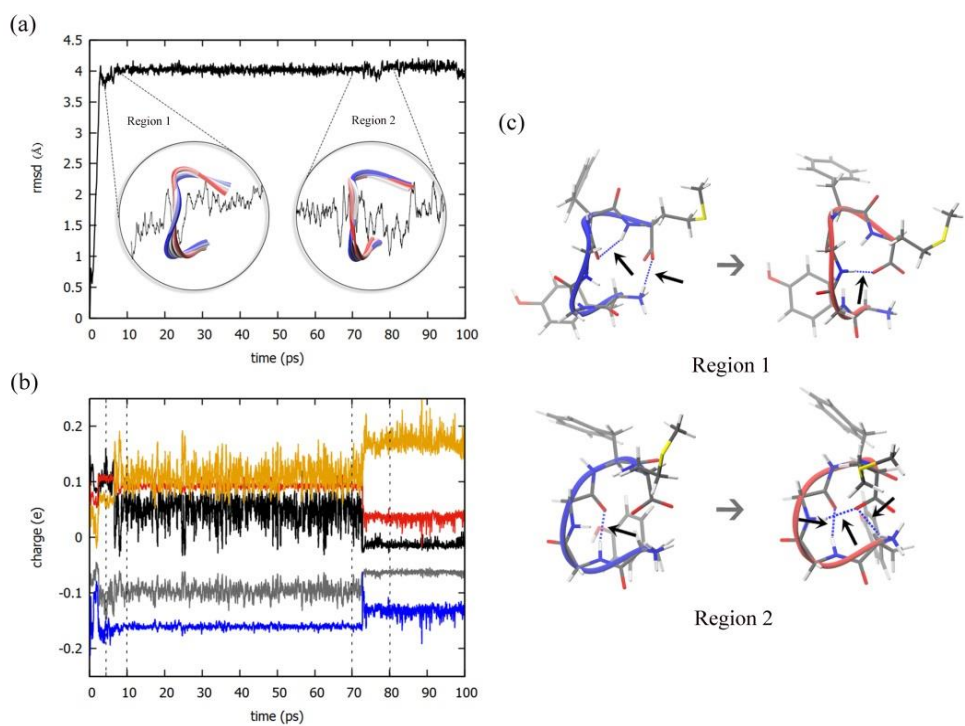

**Figure 8**. (a) The RMSD for the charged MetEnk system for a specific trajectory, using the PEM@AMBER force field. Insets are the overlapped peptide backbone. (b) The time evolution of residue charges population for the charged MetEnk system. Note, the residue charge of Tyr-1 and Met-5 is shifted to near zero by plus ±1.0 e. (c) The hydrogen bond details at the Region 1 and 2 are shown.

Next, a global analysis of the hydrogen bond distribution for the folded MetEnk system is performed, since the hydrogen bond is a critical component to maintain the peptide folded states. The occupancy percentage of hydrogen bonds averaged over 50 trajectories is determined, in order to create a global view of the hydrogen bonds. After the occupancy percentage of hydrogen bonds is calculated, we use a simple network model to represent the hydrogen bond connection among residues (Figure 9). This hydrogen bond network at residue level can be drawn as a graph. Each residue is represented by a vertex. And the edge represents the hydrogen bond relationship among residues. The occupancy percentage larger than 10% is used to build the hydrogen bond network.

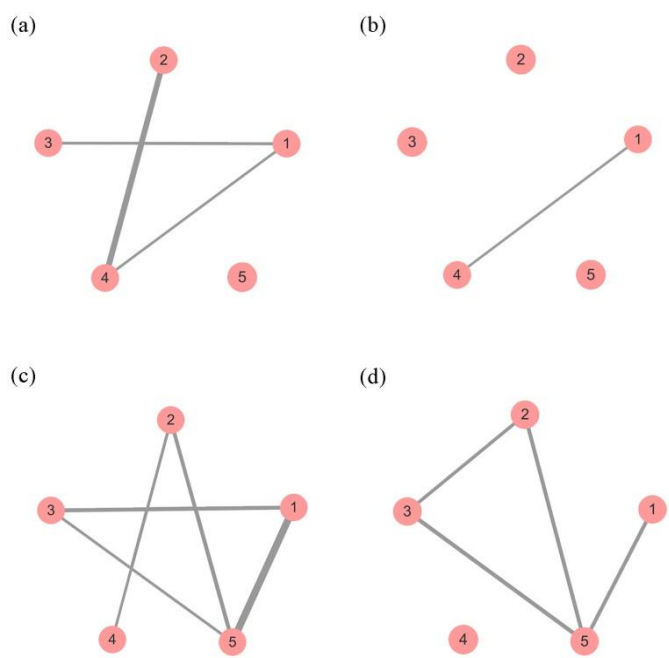

**Figure 9.** The network of MetEnk system (YGGFM) formed by connecting residues with most significant hydrogen bonds. (a) The neutral MetEnk system with AMBER scheme; (b) the neutral MetEnk system with PEM@AMBER scheme; (c) the charged MetEnk system with AMBER scheme; (d) The charged MetEnk system with PEM@AMBER scheme. The width of the edge reflects the relative important of each hydrogen bond.

As shown in Figure 9, the dominated hydrogen bonds in the network are relatively reduced in the PEM@AMBER force field. This is because the hydrogen bonds are much flexible in the charge fluctuation model and most hydrogen bonds would exchange with each other and be averaged out. Generally, the most dominated hydrogen bonds at residue level are retained. However, most significant hydrogen bonds among residues change their atomic details. Take the charged MetEnk system as an example, the hydrogen bonds between residue Phe-4 and Gly-2 is the most significant in AMBER force field, while the hydrogen bonds between Phe-4 and Tyr-1 become remarkable in PEM@AMBER force field. Therefore, the AMBER force field provides a very good global view of hydrogen bonds, at least at residue based resolution. And the PEM@AMBER force field introduces extra flexibility for hydrogen bonds in classical MD simulations.

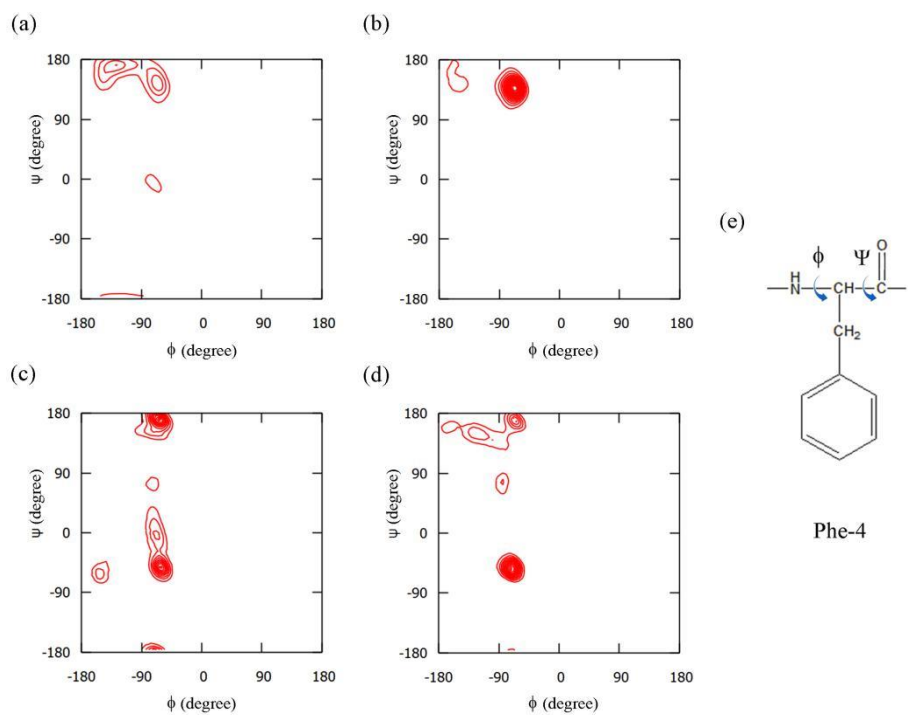

**Figure 10**. The Ramachandran plots for the Phe-4 of MetEnk systems. (a) the neutral system with AMBER scheme; (b) the neutral system with PEM@AMBER scheme; (c) the charged system with AMBER scheme; (d) the charged system with PEM@AMBER scheme.

Finally, the backbone conformation is analyzed for MD trajectories. Two torsional angles in the polypeptide chain, also called Ramachandran plot, is used, which describes the rotations of the polypeptide backbone around the bonds between N-Cα (φ) and Cα-C (ψ).[47, 65-70] Figure 10 shows the φ/ψ distribution for the conformation space sampled using AMBER and PEM@AMBER scheme, for which 50 folded trajectories were averaged. Thus, we only focus on the φ/ψ distribution of Phe-4 residue in our discussions.

The dedicated details of the distinct regions are not completely the same. As shown in Figure 10b, the observed clustering of charged MetEnk system at ψ = 180° and ψ = 0° is weaken, meanwhile the clustering near ψ = -90° is enhanced, with the charge fluctuation PEM@AMBER model. For the neutral system, the distinct regions are evenly reduced and concentrated near ψ = 180°. These features may be explained in terms of the backbone dipole-dipole interactions[17, 71-72], which can be modified by the charge flux effects in PEM@AMBER scheme. Thus, we suggest the charge fluctuation force field based on PEM@AMBER scheme enhanced the rigidity of backbone conformations in contrast to the classical AMBER force field.

In summary, the global view of the φ/ψ distribution of Phe-4 shows consistent between AMBER and PEM@AMBER scheme, which means the PEM@AMBER did not discover any new distinct φ/ψ region. Only the weights of distinct regions are slightly modified in different force fields. This also indicates the clustering of dynamics trajectories from AMBER force field provides good rough initial data sets for developing PEM@AMBER scheme.[73]

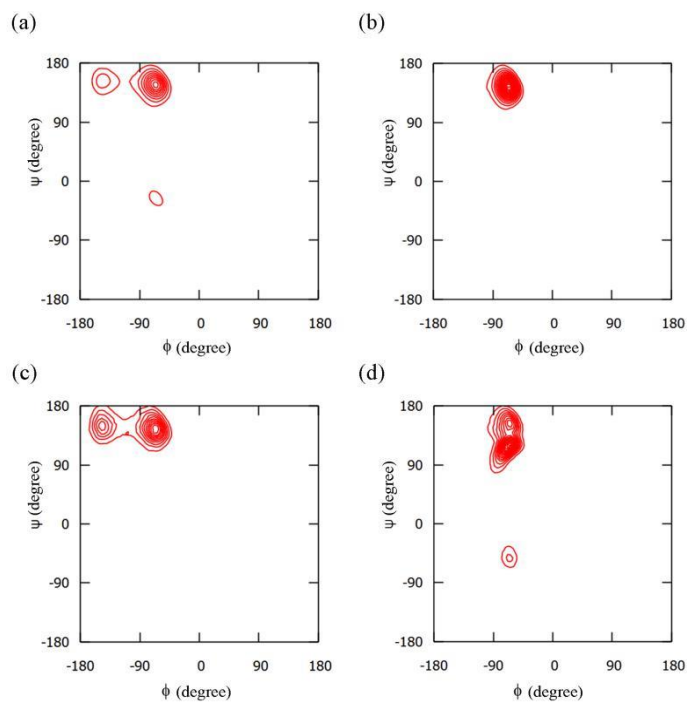

**Figure 11**. The Ramachandran plots for the Phe-4 of MetEnk systems in implicit solvent. (a) the neutral system with AMBER scheme; (b) the neutral system with PEM@AMBER scheme; (c) the charged system with AMBER scheme; (d) the charged system with PEM@AMBER scheme.

We also performed MD simulation in the present of implicit solvent. And the GBIS model is applied to for both neutral and charged MetEnk systems. Figure 11 shows the φ/ψ torsional distribution. In contrast to the gas phase results, the flexibility of the backbone motion is relatively prohibited. Similarly, the PEM@AMBER force field did not observed any new and significant distinct region in the Ramachandran plots. Only a minor distinct region between ψ = -90 ° and ψ = 0 ° is observed, for the charged MetEnk system. And the distribution of the backbone conformation seems to be concentrated in PEM@AMBER force field than AMBER force field with fixed point charge model.

## Discussions

The explicit inclusion of polarization and charge transfer in empirical force field is a long-standing objective of molecular modeling. The increasing computational power enhances our ability to build even big data for model development. By using carefully created benchmark data as the starting point, nonlinear associations between atomic configurations and molecular properties may be learned by induction. We proposed a data driven model, i.e. D$^2$FF/PEM, as an alternative choice to take into account of the charge flux effects in the oligopeptides.

In the D$^2$FF/PEM scheme, the clustering analysis of heterogeneous ensembles for MetEnk is applied to identify the truly significant representations, which can be used in a non-linear combination to describe any other structure required. Then, the ensemble averaging creates a group of coarse grained patterns, each with low bias and high variance, and combines them to represent entire structure space with low bias and low variance. It is thus a possible solution of the bias-variance dilemma. This scheme is directly applied to update atomic charges in classical MD simulations on the top of AMBER force field, namely PEM@AMBER.

The charge fluctuation PEM@AMBER force field and AMBER *ff14SB* force field provide similar global view of the peptide folding process, although the charge fluctuation speeds up the folding process in the case of the MetEnk system.[61-64] In contrast to the fixed point charge AMBER force field, the PEM@AMBER scheme provides flexible hydrogen bonds among residues; meanwhile, backbone conformational rigidity is relatively enhanced as shown in the Ramachandran plots. In addition, the significant charge flux events are directly "encoded" in the PEM@AMBER scheme, which may be helpful to understand structure dependent partial charges. The atomic charge variation can be directly observed, while it cannot happen with the fixed point charge force field.

**Conclusions**

In summary, the data driven approach provides us an alternative route toward steadily improving the traditional force field in the foreseen big data scenario. The D$^2$FF/PEM scheme could incorporate the charge flux effects on the basis of any fixed point charge force field with minimal modification of the potential function forms, which can be easily coded in available MD packages. This D$^2$FF/PEM scheme is bootstrapping in nature, which refers to a self-starting process that is only automatically trained on the MD trajectory itself. Thus, the obtained D$^2$FF/PEM scheme can be re-optimized by using the generated MD trajectories iteratively. Further work is going on to study large biomolecules, by combining the molecular fragments and transfer learning algorithms.

## Acknowledgement

The work is supported by National Natural Science Foundation of China (Nos. 21503249, 21373124), and Huazhong Agricultural University Scientific & Technological Self-innovation Foundation (Program No.2015RC008), and Project 2662016QD011 and 2662015PY113 Supported by the Fundamental Founds for the Central Universities. The authors also thank the support of Special Program for Applied Research on Super Computation of the NSFC-Guangdong Joint Fund (the second phase) under Grant No.U1501501.